\documentclass{emulateapj}

\slugcomment{Pre-print version,  Astrophysical Journal letters; received 2006 January 6; 
accepted 2006  March 2}

\shorttitle{Far--IR excited OH lines from Orion KL outflows}
\shortauthors{Goicoechea et al.}

\begin{document}

\title{Far--IR excited OH lines
from Orion~KL outflows\thanks{Based on observations with ISO,
an ESA project with instruments funded by ESA Member States
(especially the PI countries: France, Germany, the Netherlands
and the United Kingdom) and with participation of ISAS and NASA.}}

\author{Javier R. Goicoechea\altaffilmark{2}, 
Jos\'e Cernicharo\altaffilmark{3}, Mercedes R. Lerate\altaffilmark{4,5},
Fabien Daniel\altaffilmark{3}, \\ Michael J. Barlow\altaffilmark{5}, 
Bruce M. Swinyard\altaffilmark{4}, Tanya L. Lim\altaffilmark{4}, Serena 
Viti\altaffilmark{5}, Jeremy Yates\altaffilmark{5}}

\altaffiltext{2}{LERMA, UMR 8112, CNRS, Observatoire de Paris 
and \'Ecole Normale Sup\'erieure, 24 Rue Lhomond, 75231 Paris 05, 
France. (javier@lra.ens.fr)}

\altaffiltext{3}{DAMIR, Instituto de Estructura de la Materia, Consejo
Superior de Investigaciones Cient\'{i}ficas, 
Serrano 121, 28006, Madrid, Spain. (cerni,daniel@damir.iem.csic.es)}

\altaffiltext{4}{Rutherford Appleton Laboratory, Chilton, 
Didcot OX11 0QX, UK. (M.R.Lerate,B.M.Swinyard,T.L.Lim@rl.ac.uk)}

\altaffiltext{5}{University College London, Gower Street, 
London WC13, UK. (mjb,sv,jyates@star.ucl.ac.uk)}

\begin{abstract}

As  part of the first far-IR line survey towards Orion~KL, we present
the detection 
of seven new rotationally excited  OH $\Lambda$--doublets 
(at $\sim$48, $\sim$65, $\sim$71, $\sim$79, $\sim$98 and $\sim$115~$\mu$m). 
Observations were  performed  with the \textit{Long Wavelength Spectrometer}
(LWS) Fabry--Perots on board the \textit{Infrared Space Observatory} (ISO).
In total, more than 20 resolved OH rotational lines,
with upper energy levels up to $\sim$620~K, have been detected at an
angular and velocity resolutions of $\sim$80$''$ and $\sim$33~km~s$^{-1}$ 
respectively. OH line profiles show a complex behavior evolving from 
pure absorption, P--Cygni type to pure emission. 
We also present a large scale
6$'$ declination raster in the OH  $^2\Pi_{3/2}$ $J$=5/2$^+$--3/2$^-$
and $^2\Pi_{3/2}$ $J$=7/2$^-$--5/2$^+$ lines (at 119.441 and 84.597~$\mu$m)
revealing the decrease of excitation outside the core of the cloud.
From the  observed profiles, mean intrinsic line widths 
 and velocity offsets between emission and absorption line peaks 
we conclude that most of the excited OH arises from Orion outflow(s), 
i.e. the  \textit{``plateau''} spectral component.
We determine an averaged OH abundance relative to H$_2$
of $\chi(OH)$=(0.5--1.0)$\times$10$^{-6}$, a kinetic temperature of 
$\gtrsim$100~K  and a density of $n(H_2)$$\simeq$5$\times$10$^5$~cm$^{-3}$. 
Even with these conditions, the OH excitation is heavily coupled
with the strong dust continuum emission from the inner \textit{"hot
core"} regions and from the expanding flow itself.

\end{abstract}


\section{Introduction}

The Kleinmann--Low (KL) infrared (IR; L$\sim$10$^5$~L$_\odot$) nebula 
in the Orion molecular cloud is the nearest ($\sim$450~pc) and probably
the most studied massive star forming region \citep{gen89}. 
Early studies soon realized that  the large scale  distribution of gas and dust was heavily influenced by violent phenomena such as the interaction of compact and large scale 
outflows with the quiescent gas  producing strong line and continuum emission.
 IRc2 was believed to be the main source of luminosity,  heating and dynamics of the region.
However, the great advances of near-- and mid--IR 
subarcsecond resolution imaging and of (sub)millimeter interferometric
observations have dramatically changed our understanding
of the region. 
First, the 8--12~$\mu$m emission peak of IRc2 is not
coincident with the Orion SiO maser origin (related to the outflow(s) origin)
and second, its intrinsic IR luminosity (L$\sim$1000~L$_\odot$) is only a fraction of the complex luminosity \citep{gez98}. In addition,  3.6--22~$\mu$m images show that IRc2 is in fact resolved into four components that may even not be self--luminous.  
Therefore, the relevance of Irc2 as the powerful
engine of  Orion~KL is not longer supported and its nature
is now even less clear \citep{dou93,shu04,gre04}. 
A new step forward was given by \citet{men95} with
the detection of the very embedded radio continuum source 
$I$ (located 0$''$.5 south of IRc2) 
as the source that coincides with the centroid 
of the SiO maser distribution. 
\citet{men95} also detected the radio continuum emission
of IR source $n$, and suggested that it could also contribute
to the origin of some of the phenomena observed at larger scales.
Thus, in addition to $BN$, the core of Orion~KL contains at least
two more compact HII regions ($I$ and $n$) that seem 
to be running away from a common point, suggesting
that $BN$, $I$ and $n$ were originally part of a common massive stellar
system that disintegrated $\sim$500 years ago \citep{gom05}.
A causal relation between the dynamical decay of this stellar  
system  and the observed large scale molecular outflow(s), H$_2$ fingers
and bow shocks \citep{sto98} has been proposed  \citep{bal05}.
\begin{figure*} [ht] 
\centering
\includegraphics[angle=-90,width=16.cm]{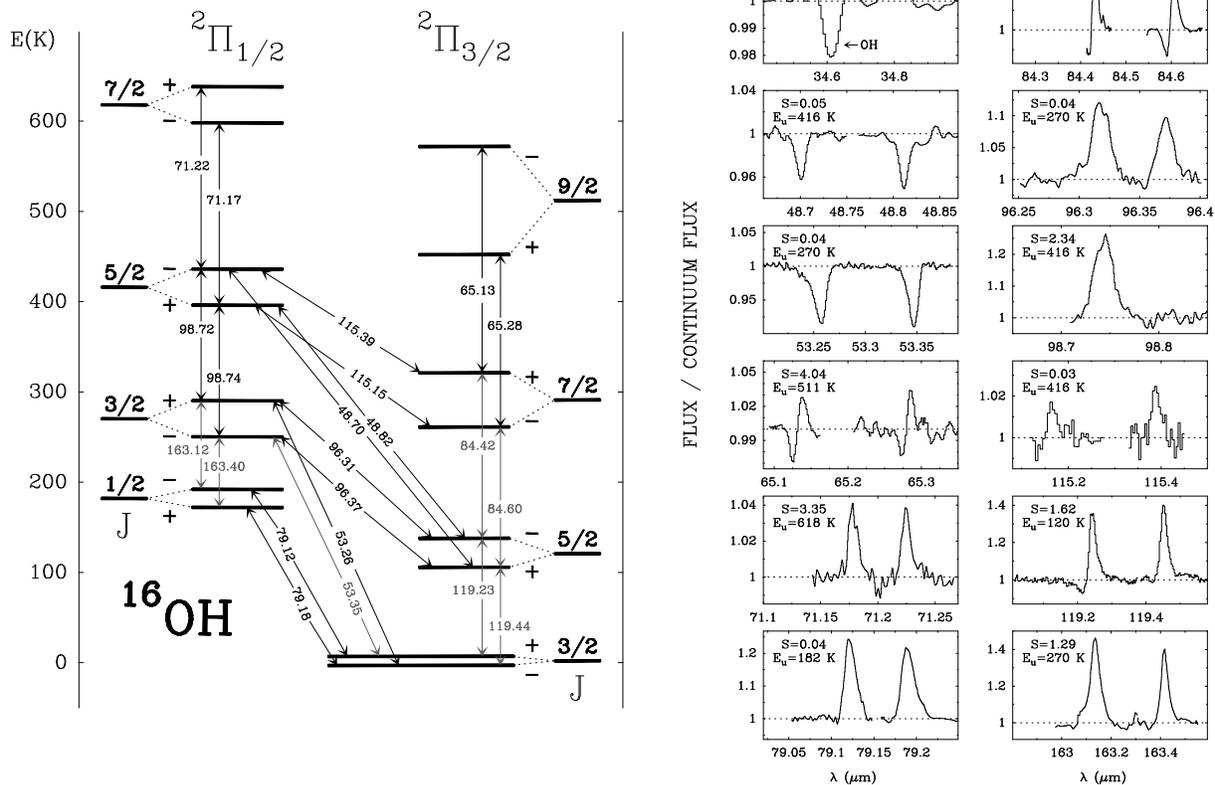}
\caption{\textit{Left:} Rotational energy diagram of $^{16}$OH showing
    the lines detected by ISO in $\mu$m. Newly detected
    transitions are shown in black, while previous KAO detections are shown
    in grey. The $^2\Pi_{3/2}$ and $^2\Pi_{1/2}$ rotational ladders 
    produced by the spin--orbit interaction are shown. The $\Lambda$--doubling
    splitting  of each rotational level has been enhanced for clarity. 
    Hyperfine structure  is not shown.
    \textit{Right:} ISO-LWS-FP observations of $^{16}$OH 
    towards Orion. The ordinate corresponds to the continuum normalized 
    flux and the abscissa to the wavelength in $\mu$m. In each box, the 
    upper level  energy (in~K) and the intrinsic line strength of the 
    transition are also 
    shown. The first box corresponds to the not resolved OH 
    $^2\Pi_{3/2}$--$^2\Pi_{1/2}$ $J$=3/2--5/2 lines at $\sim$34~$\mu$m
     detected by  the ISO-SWS \citep{wri00}.}
\label{fig-observations}
\end{figure*}

The different physical conditions and velocity fields along the line of sight 
result in a segregation of the gas and dust chemistry
that complicates the interpretation of observations toward Orion, especially
the low angular resolution molecular surveys. 
For this reason, it is common to distinguish between different spectral
components: 
the \textit{``ridge''}  (extended quiescent molecular gas), 
the \textit{``hot core''} (collection
of very dense and hot gas clumps in which sources $I$ and $n$ are embedded)
and the \textit{``plateau''} (a mixture of outflow(s), shocks and interactions
with the ambient cloud).
Submm aperture synthesis line surveys have finally provided the  spatial location
and  extent of many  molecular species
\citep{bla96,wri96,liu02,beu05}. As a result, the formation
of complex oxygen--bearing species 
in the interaction region between
the outflow(s) and the quiescent gas is now generally understood in 
the context of an oxygen--rich molecular outflow
\citep{bla87,wri96,liu02}. However, the thermal lines from basic oxygen 
reservoirs (O, H$_2$O and OH) lie in the far--IR,
a spectral region blocked by the earth atmosphere. 
Besides, the
wide range of excitation conditions provided by oxygen hydrides  supply
unique information about the energetics and dynamics of the region.
In particular,  far--IR OH line emission (first detected by
\citet{sto81}) is predicted to be a powerful
diagnostic of  the \textit{plateau} gas  \citep{dra82}.
Observations with the {\it Kuiper Airborne Observatory (KAO)}
allowed the detection of several low excitation OH rotational lines,
see Fig.~\ref{fig-observations} \citep{wat85,vis85,bet89,mel87,mel90}. 
The specific contribution of the \textit{hot core}, \textit{plateau} and
\textit{ridge} components and the physical conditions within the OH 
emitting gas remains questionable.

\section{Observations and Data reduction}

\begin{figure} [ht] 
\centering
\includegraphics[angle=0,width=7.2cm]{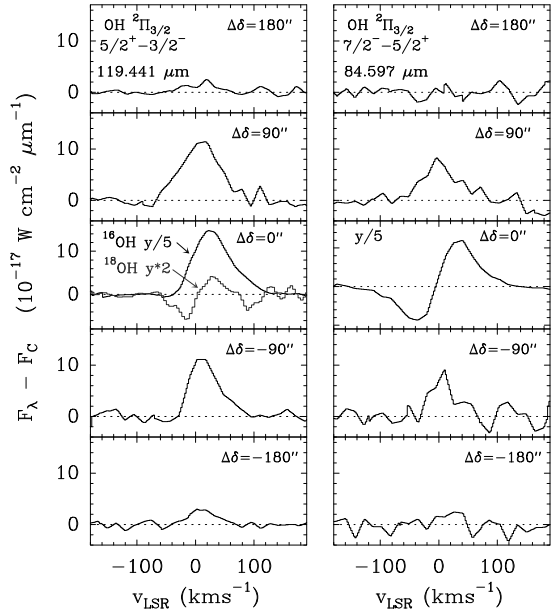}
\caption{Continuum subtracted ISO--LWS--FP spectra taken at different 
declination offsets from Orion~IRc2. Offsets in arcsec are indicated.
The left (0$''$,0$''$) panel shows the different 
ground--state line profiles produced by $^{16}$OH and 
$^{18}$OH (also detected by KAO).}
\label{fig-raster}
\end{figure}

Most of the OH pure rotational lines appear in the wavelength range of the
ISO/LWS instrument \citep{cle96}.
The LWS circular aperture size is $\sim$80$''$.
In its Fabry--Perot mode (FP)
the instrumental response is close to a Lorentzian
with a spectral  resolution of $\sim$33~km~s$^{-1}$.
In this letter we present observations that are  part
of the first full far--IR line survey of Orion~KL using the 
unprecedent wavelength coverage
and velocity resolution of the ISO-LWS-FP 
(Lerate et al. 2006, in prep.). OH spectra
were obtained in the  \textit{Astronomical Observation Template} (AOT)
L04 and L03 modes.
Processing of the OH spectra from  AOT L03 was carried out using the 
Offline Processing (OLP) pipeline and the LWS Interactive Analysis (LIA)
package version 10. 
AOT L04 spectra were analyzed using the ISO Spectral Analysis Package 
(ISAP\footnote{ISAP is a joint development by the LWS and SWS Instruments
Teams and Data Centers. Constributing institutes are CESR, IAS,
IPAC, MPE, RAL and SRON.}).
Typical routines include dark current optimization, 
deglitching and removing of the LWS grating profile, oversampling and 
averaging individual scans, baseline removal and line flux measurements.
The full description
of the complex data calibration and reduction process and  associated 
target dedicated time numbers (TDTs) of the survey are given in 
Lerate et al. (2006, in prep.). The resulting OH lines are shown in 
Fig.~\ref{fig-observations}.
Finally, we present a  $\pm$180$''$ declination raster 
(see Fig.~\ref{fig-raster})  in the OH  $^2\Pi_{3/2}$ 
$J$=5/2$^+$--3/2$^-$ and $^2\Pi_{3/2}$ 
$J$=7/2$^-$--5/2$^+$ lines at 119.441 and 84.597~$\mu$m (L04: TDT 70101216).

\section{Results and Discussion}

Far--IR OH lines show a complicated behavior  evolving
from pure absorption to pure emission profiles.
The detection of  P-Cygni type profiles in 
$^2\Pi_{3/2}$ OH excited rotational
lines  at $\sim$65 and $\sim$84~$\mu$m (Fig.~\ref{fig-observations})
and in the $^{18}$OH ground--state line 
(Fig.~\ref{fig-raster}; also detected by \textit{KAO}; Melnick et al. 1990)
confirms that most of the OH emission arises from
an extended molecular outflow. Since its  optical depth is small, 
the $^{18}$OH emission helps to trace the dominant outflow 
origin of the OH emission.
Compared  with stellar wind profiles, 
the molecular  far--IR P--Cygni type profiles are favored by  the presence of
far--IR (dust) continuum emission  throughout the expanding flow 
(and not only from a central continuum source). 
Hence, the detailed balance between emission and absorption components
critically depends on the dust and OH  emissivities,
opacities  and beam filling factors at each wavelength.
Simple nonlocal non--LTE radiative transfer models for dust and OH easily reproduce 
these characteristics (see below).
On the other hand, all 
observed OH lines within the $^2\Pi_{1/2}$ ladder 
($\sim$71, $\sim$98  and $\sim$163~$\mu$m) appear in emission
with a peak velocity of $v\gtrsim +15$~km~s$^{-1}$ 
showing less indication of self--absorption
(the cloud rest velocity is $\sim$9~km~s$^{-1}$; \citet{sco83}). 
Taking into account the high energy of the associated 
levels, effective radiative pumping by 
the far--IR dust continuum emission must be playing a role in the OH excitation.
This is clearly favored by the detection of several  
$^2\Pi_{3/2}$--$^2\Pi_{1/2}$ cross--ladder transitions in almost pure
absorption ($\sim$34, $\sim$48  and $\sim$53~$\mu$m).
Hence, temperatures and densities in the outflow  
can be significantly lower than those required to thermalise the
OH emission lines.
Other newly detected OH excited cross--ladder transitions 
($\sim$96  and $\sim$115~$\mu$m) are observed in pure emission and they
 contribute to the radiative de--excitation of 
associated $^2\Pi_{1/2}$ upper levels. 
Note that the absorption peak of pure absorption OH  lines occurs at negative
velocities,  $v\lesssim -15$~km~s$^{-1}$.
Similar conclusions apply for   water  lines  \citep{cer99a}. In particular,  pure absorption 
mid--IR  H$_2$O  lines  peak at $-$8$\pm$3~km~s$^{-1}$
 \citep{wri00}. 
Taking into account the
velocity resolution and wavelength calibration of the LWS/FP instrument, 
we find 25$\pm$5~km~s$^{-1}$ as the most likely expansion
velocity of the far--IR OH flow. The inferred expansion velocity
seems more consistent with the 18$\pm$2~km~s$^{-1}$
\textit{low--velocity outflow}, originally  revealed  by water  
maser motions \citep{gen81}. 
In addition, it is known that  OH 1.6~GHz masers toward the Orion~KL region are detected over an area of 30$''$ in diameter. The best kinematical model fitting the OH maser emission  is found for an uniform expansion velocity of 21.0$\pm$3.5~km~s$^{-1}$ away from a source (at or near IRc2)
with a radial velocity of 9.0$\pm$0.5~km~s$^{-1}$ \citep{coh06}.
Contribution from the extended
\textit{high--velocity outflow} \citep{mar90} could 
also be present (see predictions by  Melnick et al. 1990). 
Only the OH ground--state lines at
$\sim$119~$\mu$m show high--velocity broad--red--wing emission above 
$\sim$100~km~s$^{-1}$. This high--velocity emission is not observed 
in any other OH line by ISO. 

The large--scale OH declination raster clearly 
shows that OH lines are weaker (at least by a factor of 5) outside the 
central position (Fig.~\ref{fig-raster}).
This is likely due to lower excitation (continuum emission and densities
decrease) in the extended cloud, but not necessarily to a steep 
decrease of OH abundance in the \textit{ridge}. 
Unfortunately, it is impossible to infer the exact spatial distribution 
of the newly detected OH lines from the large ISO beam. 
KAO heterodyne observations of the $\sim$119.2~$\mu$m line 
(with a beam of $\sim$33$''$) showed that OH emission may 
be spatially compact. \citet{bet89} deduced that the OH emission comes 
from a source $<$25$''$ in full--width at half--maximum diameter. 
Therefore, the same 
inner regions of the large scale outflow  revealed by 
H$_2$O masers at $\sim$325~GHz  \citep{cer99b}, and 
HDO  emission at $\sim$893 and  $\sim$850~GHz \citep{par01}.
Besides, OH may also arise from the regions where the outflow plunge
into the quiescent cloud producing dissociative shocks.
According to recent [C\,{\sc i}] observations (with 10$''$ angular 
resolution), molecular dissociation occurs in a shell of $\sim$40$''$ diameter 
\citep{par05},  therefore within the LWS beam.
Nevertheless, the OH emission measured by \citet{bet89} towards the bright
H$_2$ shocked regions Peak~1/2 is significantly weaker than towards
the outflow itself. More quantitative conclusions about
the spatial distribution  of the OH flow require much higher 
angular resolution  observations.\\


\begin{figure} [ht] 
\centering
\includegraphics[angle=0,width=7.5cm]{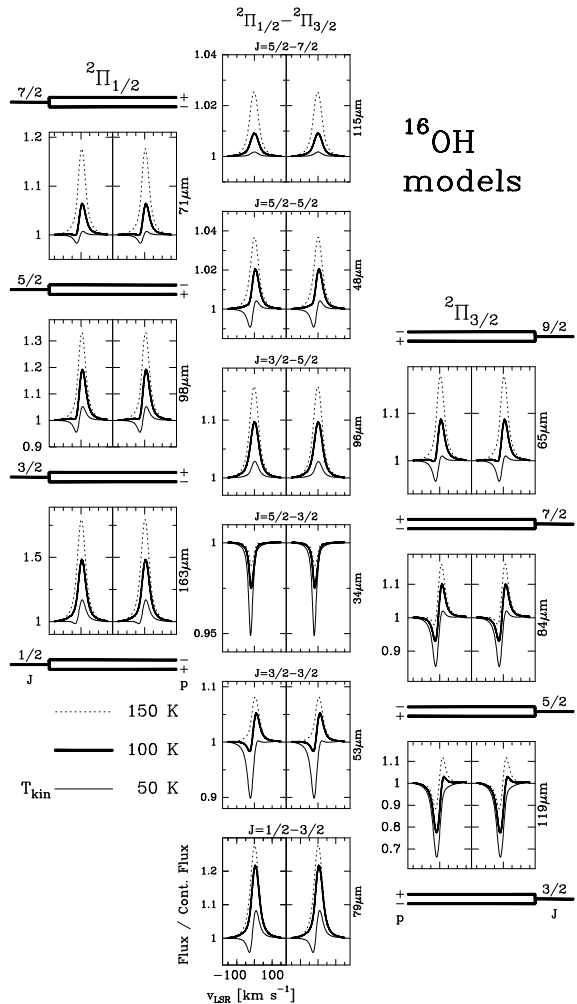}
\caption{Radiative transfer models for 
an expanding OH shell. 
Common  parameters are: $\chi$(OH)=5$\times$10$^{-7}$ and 
$n(H_2)$=5$\times$10$^5$~cm$^{-3}$. Three different  temperatures
are considered:  T$_k$ = 50, 100, and 150~K.
Each panel correspond to a rotational $\Lambda$--doublet. The wavelength
of each transition is shown. Line profiles have been convolved with 
a gaussian beam of 80$''$  and with a spectral resolution 
 Lorentzian  width of 33~km~s$^{-1}$. }
\label{fig-models}
\end{figure}

In order to estimate the OH abundance and the physical conditions leading
to the observed line profiles,
we have modeled the first 20 OH rotational levels with  the
nonlocal  code used in our OH study of  Sgr~B2  \citep{goi02}. 
We have modeled a spherical shell 
with a 25$''$ diameter expanding at 25~km~s$^{-1}$ that surrounds an 
uniform core, optically thick in the far--IR, with a 10$''$ diameter 
and a color temperature of 150~K. 
These values agree with those obtained from
high dipole molecules towards the \textit{hot core}.
In all models, the dust emission reasonably fits
the far--IR continuum levels measured by the LWS. 
Obviously this model simplifies the inhomogeneous and clumpy
nature of the region as ISO observations average all the 
sub--structure morphology and the different physical conditions found in  the \textit{plateau}.
In our view, the far--IR OH excited emission  
may be spatially associated with the SO and SO$_2$ shell of low--velocity
expanding gas, observed at $\sim$1$''$ resolution,
where the outflow shocks dense clumps of ambient  material \citep{wri96}.
 
A grid of models has been generated  by varying 
T$_{k}$=T$_{dust}$  from 50 to 250~K, $n$(H$_2$) from 
10$^5$ to 10$^7$~cm$^{-3}$ and $\chi$(OH) from 10$^{-8}$ to 
10$^{-5}$ in the outflow. The newly detected
OH excited lines add some important constraints into the \textit{plateau}
physical conditions. Basically, the fact that excitation temperatures
in several excited OH cross--ladder transitions 
have to be below  T$_{dust}$ constraints the maximum T$_k$
and $n$(H$_2$) allowed to produce absorption in these transitions. 
These absorptions can only be reproduced if 
$n$(H$_2$)$<$5$\times$10$^6$~cm$^{-3}$, otherwise collisional excitation
and re--emission dominates.
For this range of $n$(H$_2$), the bulk of gas has to
be above T$_k$=50~K, otherwise absorption will dominate the lower energy 
cross--ladder and $^2\Pi_{3/2}$ intra--ladder transitions. On the other
hand, collisional excitation will dominate for T$_k$$>$200~K.
Since high T$_k$ and $n$(H$_2$) conditions result in a OH pure emission 
line spectrum, a dominant contribution from the \textit{hot core} is not 
expected. In addition, the large far--IR line+continuum opacity in the 
\textit{plateau} itself will hide most of the \textit{hot core} OH 
emission.
The lack of spectral resolution makes a more accurate analysis of
possible absorptions impossible.

Nevertheless, the detected OH self--absorptions and P--Cygni type profiles
allows one to constraint the  physical parameters leading 
to the observed balance between collisions
and radiative pumping.
The best single--component fit to the OH observations is obtained around 
T$_k$$\sim$100~K and $n$(H$_2$)$\sim$5$\times$10$^5$~cm$^{-3}$
(see models in Fig.~\ref{fig-models}).
In comparison with far--IR H$_2$O ground--state or CO high--$J$ lines,
 OH cross--ladder transitions have both small spontaneous
emission rates and small line strengths.
In most applications this indicates that the associated far--IR
lines are optically thin. 
The lack of opacity broadening also implies that their line profiles are 
more sensitive to gas velocity fields and turbulence.
A good agreement with 
observations is found for $\chi$(OH)=(0.5--1.0)$\times$10$^{-6}$.
Therefore, \textit{ISO} observations clearly
show that OH is very abundant in the  \textit{plateau}. 
These results provide additional insights to the fact that
complex O--rich molecules 
such as methanol (CH$_3$OH) \citep{wri96} or formid acid (HCOOH) \citep{liu02}
are  specifically enhanced (when resolved at much higher resolution) 
in the interaction surfaces between the outflow (OH rich) material  
and the ambient cloud. 
Only the wide range of excitations 
provided by water lines 
and the input from chemical models, able to predict the chemistry evolution
of the physically distinct regions in Orion,
can extract  more accurate information from far--IR observations.

\acknowledgments

We are grateful to the LWS instrument and the ISO
data base teams for the quality of the provided data.
The referee help us to improve our high resolution view of Orion.
JRG was supported by a \textit{Marie Curie Intra-European Individual
Fellowship} within the 6th European Community Framework Programme,
contract MEIF--CT--2005--515340.
We  thank the Spanish DGES and PNIE grants 
ESP2001-4516 and AYA2003-2785.


\end{document}